\documentclass[aps,prb,reprint,twocolumn,superscriptaddress,showpacs,floatfix,longbibliography]{revtex4-1}

\usepackage{amsmath}
\usepackage{color}
\usepackage{mathrsfs}
\usepackage{dcolumn}
\usepackage{bm}
\usepackage{multirow}
\usepackage{graphicx}
\usepackage{rotating}
\usepackage{layout}
\usepackage[version=3]{mhchem}
\usepackage{braket}
\usepackage{longtable}
\usepackage{setspace}

\allowdisplaybreaks
\usepackage{xfrac}

\usepackage[colorlinks,citecolor=blue,urlcolor=blue,bookmarks=false,hypertexnames=true]{hyperref}

\begin{document}

\title{Accurate prediction of inverted singlet-triplet excited states using self-consistent spin-opposite perturbation theory}
% Force line breaks with \\
\author{Nhan Tri Tran}
\email{trantrinhan@vlu.edu.vn}
\affiliation{Simulation in Materials Science Research Group, Science and Technology Advanced Institute, Van Lang University, Ho Chi Minh City, Vietnam}
\affiliation{Faculty of Applied Technology, Van Lang School of Technology, Van Lang University, Ho Chi Minh City, Vietnam}

\author{Hoang Thanh Nguyen}
\affiliation{Institute of Advanced Technology, Vietnam Academy of Science and Technology, Ho Chi Minh City 700000, Vietnam}

\author{Lan Nguyen Tran}
\email{tnlan@hcmus.edu.vn}
\affiliation{University of Science, Vietnam National University, Ho Chi Minh City 700000, Vietnam}
\affiliation{Vietnam National University, Ho Chi Minh City 70000, Vietnam}

\begin{abstract}

The violation of Hund’s rule, resulting in an inverted singlet-triplet (INVEST) gap, represents a paradigm shift in photophysics with major implications for OLED technology. INVEST molecules facilitate barrierless reverse intersystem crossing, theoretically permitting 100\% internal quantum efficiency without thermal activation. However, accurately predicting negative singlet-triplet energy gaps typically demands prohibitive computational costs. In this study, we evaluate the efficacy of our recently developed one-body Møller–Plesset perturbation theory (OBMP2) and its spin-opposite variant (O2BMP2) as efficient alternatives. Benchmarking against 30 INVEST molecules reveals that O2BMP2, with appropriate spin-opposite scaling, achieves the accuracy of ADC(3) and EOM-CCSD. Furthermore, with the possibility of reducing computational complexity to $N^4$, O2BMP2 provides a robust balance of accuracy and efficiency, making it suitable for the high-throughput screening of next-generation INVEST materials.
\end{abstract}

\maketitle

\section{Introduction}
Organic Light Emitting Diodes (OLEDs) have attracted significant attention in display and lighting technologies due to their outstanding energy efficiency, high contrast ratios, and design flexibility. However, the recombination of charge carriers results in a 1:3 ratio of singlet and triplet excited states, which limits the internal quantum efficiency of OLEDs and leads to energy losses. Numerous efforts have been dedicated to improving OLED efficiency. Among the most promising strategies is the development of organic molecules with inverted singlet-triplet energy gaps (INVEST), where the first excited singlet state (S1) is energetically lower than the lowest triplet state (T1), thus breaking Hund's rule \cite{ehrmaier2019singlet,bedogni2023shining}. This energetic inversion significantly enhances reverse intersystem crossing (RISC) rates and facilitates exothermic RISC, allowing for rapid and thermally barrier-free triplet-to-singlet upconversion. Consequently, the internal quantum efficiency could potentially reach 100\%. However, experimentally quantifying the S1-T1 energy gap, $\Delta E_{\text{ST}}$, poses challenges due to the spin-forbidden nature of the S1 to T1 transitions observed in spectroscopy. As a result, quantum chemical calculations have been primarily employed to compute $\Delta E_{\text{ST}}$, thereby expanding the landscape of potential INVEST molecules \cite{wrigley2025singlet,won2023inverted}. Recently, high-throughput virtual screening, coupled with symmetry-based design strategies, has facilitated a systematic exploration of vast chemical spaces, resulting in the identification of several new candidates that exhibit inverted singlet–triplet energy gaps \cite{pollice2021organic, pollice2024rational}.

$\Delta E_{\text{ST}}$ is conventionally defined as the difference between the Hamiltonians restricted to singlet and triplet subspaces,  corresponding to twice the matrix of positive-semidefinite two-electron integrals \cite{won2023inverted}. Consequently, the inclusion of doubly excited configurations is essential for an accurate treatment of INVEST molecules \cite{drwal2023role}. As a result, widely used linear-response approaches for excited states, such as configuration interaction with single excitations (CIS) and time-dependent density functional theory (TD-DFT) under the Tamm–Dancoff approximation, fail to reliably predict $\Delta E_{\text{ST}}$. The inadequacy of TD-DFT for INVEST excited states and the significance of double excitation configurations have been examined in detail in several studies \cite{ricci2021singlet, ghosh2022origin}. Recently, it has been demonstrated that TD-DFT calculations employing frequency-dependent kernels, double-hybrid functionals, and other schemes that explicitly or implicitly incorporate double excitations can yield negative values of $\Delta E_{\text{ST}}$ \cite{bhuyan2025tuned}. Nevertheless, the predictive performance of these approaches remains inconsistent.

Delta self-consistent field methods ($\Delta$SCF) \cite{MOM-JPCA2008,STEP-JCTC2020,OODFT-JPCL2021} constitute a variational alternative to linear-response theory for the treatment of excited states. $\Delta$SCF has been extensively employed to describe X-ray absorption \cite{norman2018simulating,michelitsch2019efficient}, valence excited states, and core-level excitations \cite{Gill2009dscf,MOM-CC-2022,MOM-CC-2022,hirao2023core,jorstad2022delta}, as well as nonadiabatic molecular dynamics simulations \cite{vandaele2022deltascf,vandaele2022photodissociation,malis2021deltascf}. For INVEST molecules, several studies have computed vertical excitation energies with $\Delta$SCF \cite{valverde2025can,kunze2024deltadft}. Numerical results indicate markedly improved agreement with experimental data for both S1 and T1 energies relative to TD-DFT, implying that orbital relaxation is crucial for an explicit description of double and higher-order excitations in these emissive states. Recently, quasiparticle energy DFT (QE-DFT), wherein STGs are obtained directly from differences in Kohn–Sham orbital energies, has been shown to provide a highly efficient and accurate alternative for predicting INVEST molecules \cite{li2025efficient}. Consequently, with favorable computational cost, $\Delta$DFT is anticipated to be suitable for high-throughput virtual screening. Nevertheless, one of the principal challenges of DFT remains system dependent \cite{valverde2025can,kunze2024deltadft}. 

Correlated wavefunction methods, including second-order approximate coupled cluster (SCS-CC2) and its spin-component scaling variant (SCS-CC2), second- and third-order algebraic diagrammatic construction theory (ADC(2) and ADC(3)), equation-of-motion coupled-cluster singles and doubles (EOM-CCSD), $N$-electron valence state second-order perturbation theory (NEVPT2), and multireference second-order complete active space perturbation theory (CASPT2), have demonstrated negative $\Delta E_{\text{ST}}$ for a broad class of systems \cite{pollice2021organic,pollice2024rational,loos2023heptazine}. Notably, Loos and co-workers showed that CC2 and ADC(2) can approach CC3 and EOM-CCSDT accuracies \cite{loos2023heptazine}. Despite their accuracy, correlated wavefunction methods are prohibitively expensive for high-throughput screening of thousands or millions of molecules \cite{pang2025molecular}. Hence, there remains a strong demand for affordable yet reliable methods to predict INVEST excited states. 

We have developed a novel self-consistent perturbation theory, termed one-body M{\o}ller–Plesset perturbation theory (OBMP2) \cite{OBMP2-JCP2013,OBMP2-JPCA2021, OBMP2-PCCP2022, OBMP2-JPCA2023}. In OBMP2, a canonical transformation \cite{CT-JCP2006,CT-JCP2007,CT-ACP2007,CT-JCP2009,CT-JCP2010,CT-IRPC2010} followed by a cumulant approximation \cite{cumulant-JCP1997,cumulant-PRA1998,cumulant-CPL1998,cumulant-JCP1999} yields an effective one-body Hamiltonian. The resulting operator comprises the standard Fock term plus a one-body correlation potential that includes double-excitation MP2 amplitudes. Molecular orbitals and orbital energies are then optimized in the presence of correlation by diagonalizing a correlated Fock matrix. The self-consistency of OBMP2 mitigates issues associated with the non-iterative nature of standard MP2 for open-shell systems \cite{OBMP2-JPCA2021, OBMP2-PCCP2022} and for bond-stretching regimes \cite{OBMP2-JPCA2023}. In many cases, OBMP2 outperforms orbital optimized MP2 (OO-MP2) for open-shell systems \cite{OBMP2-PCCP2022}. We have further extended OBMP2 to a spin-opposite variant (O2BMP2) and demonstrated that OBMP2 and O2BMP2 can treat intermolecular charge-transfer excited states with accuracy comparable to high-level coupled-cluster methods \cite{OBMP2-JCP2025}. Consequently, assessing the performance of these methods for INVEST excited states is particularly timely, given the need to account for double excitations, charge transfer, and stronger electron correlation effects relative to exchange. We have carried out an assessment on the performance of our methods for 30 INVEST molecules that are taken from Refs~\citenum{loos2023heptazine} and \citenum{pollice2024rational}. We find that O2BMP2, with an appropriate spin-opposite scaling ($c_{\text{os}}$), can describe INVEST excited states with accuracy comparable to ADC(3) and EOM-CCSD. Importantly, the formal scaling of O2BMP2 can be reduced from $N^5$ to $N^4$, analogous to spin-opposite MP2 and OO-MP2\cite{O2-JCP2007}. Therefore, O2BMP2 offers high accuracy at affordable cost and is well suited for high-throughput screening aimed at identifying new molecular systems exhibiting INVEST excited states.

\section{Theory and computational details}

A comprehensive derivation of the OBMP2 theoretical framework is provided in Refs.~\citenum{OBMP2-JPCA2021,OBMP2-PCCP2022,OBMP2-JPCA2023,OBMP2-JPCA2024}; the method has been implemented within a locally modified version of the PySCF quantum chemistry package\cite{pyscf-2018}. The OBMP2 effective Hamiltonian is constructed via the canonical transformation formalism\cite{CT-JCP2006,CT-JCP2007,CT-ACP2007,CT-JCP2009,CT-JCP2010,CT-IRPC2010}:
\begin{align}
\hat{\bar{H}} = e^{\hat{A}^\dagger} \hat{H} e^{\hat{A}},
\label{Hamiltonian:ct}
\end{align}
where the molecular electronic Hamiltonian is expressed as
\begin{align}
  \hat{H} =  \sum_{pq}h^{p}_{q} \hat{a}_{p}^{q} + \tfrac{1}{2}\sum_{pqrs}g^{p r}_{q s}\hat{a}_{p r}^{q s}\label{eq:h1}.
\end{align}
In the above, the indices $\left\{p, q, r, \ldots \right\}$ denote general ($all$) spin orbitals. The one- and two-body second-quantized operators $\hat{a}_p^q$ and $\hat{a}_{pq}^{rs}$ are defined as $\hat{a}_p^q = \hat{a}^\dagger_p\hat{a}_q$ and $\hat{a}_{pq}^{rs} = \hat{a}^\dagger_p\hat{a}^\dagger_q\hat{a}_s\hat{a}_r$, respectively. The quantities $h_{pq}$ and $v_{pq}^{rs}$ represent the one- and two-electron integrals. Within the OBMP2 framework, the anti-Hermitian cluster operator $\hat{A}$ is restricted to double excitations:
\begin{align}
  \hat{A} = \hat{A}_\text{D} = \tfrac{1}{2} \sum_{ij}^{occ} \sum_{ab}^{vir} T_{ij}^{ab}(\hat{a}_{ab}^{ij} - \hat{a}_{ij}^{ab}) \,, \label{eq:op1}
\end{align}
where the amplitudes are determined at the second-order M{\o}ller--Plesset (MP2) level:
\begin{align}
  T_{i j}^{a b} =  \frac{g_{i j}^{a b} } { \epsilon_{i} + \epsilon_{j} - \epsilon_{a} - \epsilon_{b} } \,, \label{eq:amp}
\end{align}
with the indices $\left\{i, j, k, \ldots \right\}$ labeling occupied ($occ$) spin orbitals, $\left\{a, b, c, \ldots \right\}$ labeling virtual ($vir$) spin orbitals, and $\epsilon_{i}$ denoting the canonical orbital energy associated with spin-orbital $i$. Using Baker–Campbell–Hausdorff (BCH) expansion up to the second order for the effective Hamiltonian (Eq.~\ref{Hamiltonian:ct}) and employing some approximations, we arrive at the OBMP2 Hamiltonian as follows,

\begin{align}
  \hat{H}_\text{OBMP2} = \hat{H}_\text{HF} + \left[\hat{H},\hat{A}_\text{D}\right]_1 + \tfrac{1}{2}\left[\left[\hat{F},\hat{A}_\text{D}\right],\hat{A}_\text{D}\right]_1 \label{eq:h4}.
\end{align}
Where the subscript 1 on the commutators, $[\ldots]_1$, indicates that only one-body operators and scalar constants, obtained from the reduction of many-body operators via the cumulant decomposition\cite{cumulant-JCP1997,cumulant-PRA1998,cumulant-CPL1998,cumulant-JCP1999}, are retained. The operator $\hat{H}_\text{HF}$ denotes the standard Hartree--Fock Hamiltonian.  $\hat{v}_\text{OBMP2}$, composing of one- and second-order BCH terms, constitutes a correlation potential comprising exclusively one-body operators, whose explicit working expression reads
\begin{align}
\hat{v}_{\text{OBMP2}} = &\left[\hat{H},\hat{A}_\text{D}\right]_1 + \tfrac{1}{2}\left[\left[\hat{F},\hat{A}_\text{D}\right],\hat{A}_\text{D}\right]_1 \\
= &  \overline{T}_{i j}^{a b} \left[ f_{a}^{i} \,\hat{\Omega}\left( \hat{a}_{j}^{b} \right) 
  + g_{a b}^{i p} \,\hat{\Omega} \left( \hat{a}_{j}^{p} \right) - g^{a q}_{i j} \,\hat{\Omega} \left( \hat{a}^{b}_{q} \right) \right] \nonumber \\  &- 2 \overline{T}_{i j}^{a b}g^{i j}_{a b} 
   + \,f_{a}^{i}\overline{T}_{i j}^{a b}\overline{T}_{j k}^{b c} \,\hat{\Omega} \left(\hat{a}_{c}^{k} \right) \nonumber \\ 
     &+  f_{c}^{a}T_{i j}^{a b}\overline{T}_{i l}^{c b} \,\hat{\Omega} \left(\hat{a}^{l}_{j} \right) + f_{c}^{a}T_{i j}^{a b}\overline{T}_{k j}^{c b} \,\hat{\Omega} \left(\hat{a}^{k}_{i} \right) \nonumber \\ 
     &-  f^{k}_{i}T_{i j}^{a b}\overline{T}_{k l}^{a b} \,\hat{\Omega} \left(\hat{a}_{l}^{j} \right)
     -  f^{p}_{i}T_{i j}^{a b}\overline{T}_{k j}^{a b} \,\hat{\Omega} \left(\hat{a}^{p}_{k} \right) \nonumber \\ 
     & +  f^{k}_{i} T_{i j}^{a b}\overline{T}_{k j}^{a d} \,\hat{\Omega}\left(\hat{a}_{b}^{d} \right) +  f_{k}^{i}T_{i j}^{a b}\overline{T}_{k j}^{c b} \,\hat{\Omega} \left(\hat{a}_{a}^{c} \right) \nonumber \\ 
     &-  f_{c}^{a}T_{i j}^{a b}\overline{T}_{i j}^{c d} \,\hat{\Omega} \left(\hat{a}^{b}_{d} \right) \,
     - f_{p}^{a}T_{i j}^{a b}\overline{T}_{i j}^{c b} \,\hat{\Omega} \left(\hat{a}^{p}_{c} \right) \nonumber \\
     & - 2f_{a}^{c}{T}_{i j}^{a b}\overline{T}_{i j}^{c b} +  2f_{i}^{k}{T}_{i j}^{a b}\overline{T}_{k j}^{a b}. \label{eq:vobmp2} 
\end{align}
Here, $\overline{T}_{ij}^{ab} = {T}_{ij}^{ab} - {T}_{ji}^{ab}$ denotes the antisymmetrized amplitude, $\hat{\Omega} \left( \hat{a}^{p}_{q} \right) = \hat{a}^{p}_{q}  + \hat{a}^{q}_{p}$ is the symmetrization operator, and the Fock matrix elements are defined as
\begin{align}
    f_p^q = h_p^q + \sum_{i}^{occ}\left(g^{p i}_{q i} - g^{p i}_{i q} \right).
\end{align}
The OBMP2 Hamiltonian $\hat{H}_\text{OBMP2}$ (Eq.\ref{eq:h4}) can be reformulated in a manner analogous to the standard Hartree--Fock formalism:
\begin{align}
  \hat{H}_\text{OBMP2} = & \hat{\bar{F}} + \bar{C}, \label{eq:h5}
\end{align}
where $\bar{C}$ is a scalar constant aggregating all terms devoid of excitation operators. The quantity $\hat{\bar{F}}$ denotes the correlated Fock operator, $\hat{\bar{F}} =  \bar{f}^{p}_{q} \hat{a}_{p}^{q}$, with the correlated Fock matrix elements $\bar{f}^{p}_{q}$ expressed as
\begin{align}
\bar{f}^{p}_{q} &= f^{p}_{q} + v^{p}_{q}. \label{eq:corr-fock}%, \\
\end{align}
The correlation potential $v^{p}_{q}$ modifies the uncorrelated Hartree--Fock electronic structure. Consequently, the molecular orbital (MO) coefficients and orbital energies are obtained as the eigenvectors and eigenvalues, respectively, of the correlated Fock matrix $\bar{f}^{p}_{q}$. In our recent work\cite{OBMP2-JPCA2024}, we introduced spin-opposite scaling into the MP2 amplitude as
\begin{align}
  T_{i j}^{a b} =  c_{\text{os}} \frac{g_{i j}^{a b} } { \epsilon_{i} + \epsilon_{j} - \epsilon_{a} - \epsilon_{b} } \,, \label{eq:amp_os}
\end{align}
yielding the spin-opposite-scaled variant of OBMP2, designated O2BMP2. 

To selectively converge upon a specific electronic state during the OBMP2 self-consistent procedure, we utilize the maximum overlap method (MOM) algorithm, originally developed for Hartree--Fock (HF) and density functional theory (DFT) calculations\cite{MOM-JPCA2008, MOM-JCTC2018}. The procedure begins with a MOM-HF calculation targeting the desired state, the solution of which subsequently serves as the initial reference for re-optimization within the OBMP2 and O2BMP2 frameworks. Additionally, the direct inversion of the iterative subspace (DIIS) technique\cite{diis1980} is employed to accelerate convergence. 

\begin{figure*}[t!]
  \includegraphics[width=15cm]{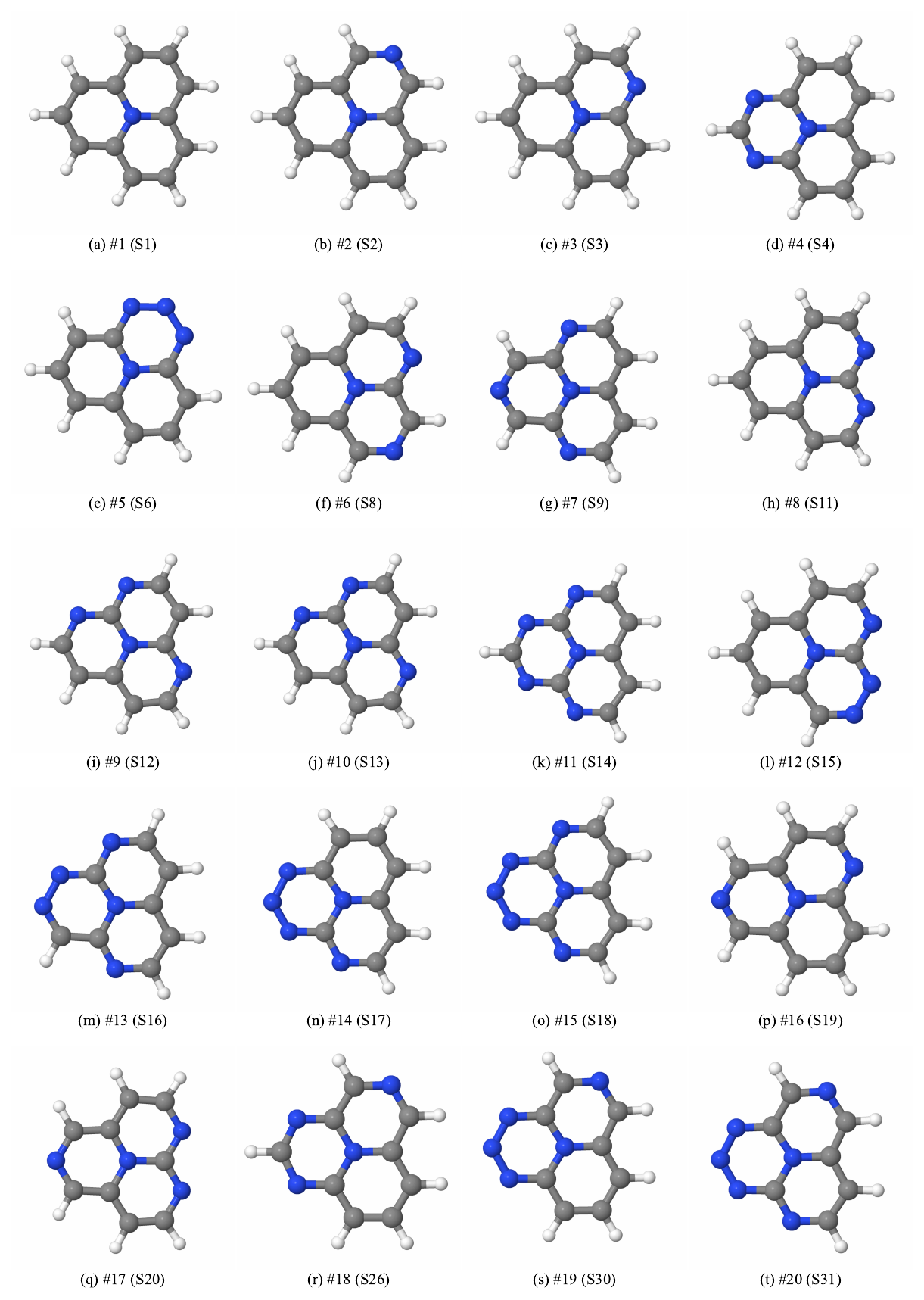}
  \caption{20 molecules in the INVEST set B adopted from Ref~\citenum{pollice2024rational}.}
  \label{fig:INVESTB}
\end{figure*}

In this work, we employ 30 molecules divided into two test sets: INVEST sets A and B. INVEST set A comprises 10 molecules taken from Ref.\citenum{loos2023heptazine}, whereas INVEST set B includes 20 molecules selected from a large dataset reported in Ref.\citenum{pollice2024rational}, as depicted in Figure~\ref{fig:INVESTB}. For comparison, we also performed $\Delta$DFT, $\Delta$HF, and $\Delta$MP2 calculations using PySCF.\cite{pyscf-2018} The aug-cc-pVDZ basis set\cite{kendall1992} was used for INVEST set A, while the cc-pVDZ basis set\cite{DUNNING:1989:ccpvdz} was used for INVEST set B. It should be noted that the formal computational scaling of OBMP2 and O2BMP2 is $\mathcal{O}(N^5)$, and convergence to a threshold of $10^{-8}$~Ha is typically achieved within several tens of iterations.

\section{Results and discussion}
\subsection{INVEST SET A}

Let us consider the INVEST set A, consisting of 10 INVEST molecules proposed by Loos and co-workers \cite{loos2023heptazine}, and examine several factors that influence the singlet–triplet energy gap, $\Delta E_{\text{ST}}$. Previous works\cite{wrigley2025singlet,won2023inverted} indicate that negative $\Delta E_{\text{ST}}$ arises from a dominance of electron correlation effects relative to exchange, implying that the spin-opposite component can significantly modify $\Delta E_{\text{ST}}$. In Table~\ref{tab:cos}, we report $\Delta E_{\text{ST}}$ values computed with OBMP2 and O2BMP2 for several spin-opposite scaling factors, namely $c_{\text{os}} = 1.2$, 1.5, and 1.8. CC2, ADC(2), and their spin-opposite variants (SOS-CC2 and SOS-ADC(2)) are also shown, as reported in Ref~\citenum{loos2023heptazine}. We can see that SOS-CC2 and SOS-ADC(2) produce more negative $\Delta E_{\text{ST}}$ than CC2 and ADC(2). While OBMP2 delivers positive $\Delta E_{\text{ST}}$ for most molecules, O2BMP2 predicts negative values for all molecules in the set. Increasing $c_{\text{os}}$ tends to produce more negative $\Delta E_{\text{ST}}$. As detailed in the Supporting Information (SI), the increase in the spin-opposite scaling coefficient reduces both singlet and triplet excitation energies due to the amplified contribution of spin-opposite correlation; however, the singlet state experiences greater stabilization than the triplet, resulting in a larger negative singlet–triplet gap.

In Table~\ref{tab:bch}, we compare $\Delta E_{\text{ST}}$ obtained with O2BMP2/1.2 using first- and second-order Baker-Campbell-Hausdorff (BCH) expansions, evaluated at the first iteration ($1^{\text{st}}$) and at full self-consistency (converged). The first BCH order at the first iteration, corresponding to SOS-MP2 \cite{OBMP2-JCP2013}, predicts all positive $\Delta E_{\text{ST}}$. Full self-consistency reduces $\Delta E_{\text{ST}}$ and yields several negative values, although not for every molecule. Negative $\Delta E_{\text{ST}}$ is only obtained after the full self-consistency of O2BMP2 consistitng of both first and second BCH orders, highlighting the central role of electron correlation and wave-function relaxation in driving the inversion of the energy gap. Overall, our O2BMP2 formalism with the spin-opposite component and full self-consistency can accurately predict INVEST excited states.

\begin{table*}[h]
    \centering 
    \caption{$\Delta E_{\text{ST}}$ for 10 INVEST molecules adapted from Ref.~\citenum{loos2023heptazine} using different methods with the formal scaling of $N^5$: OBMP2, CC2, ADC(2) and their spin-opposite component variants.}
    \label{tab:energy_methods_2}
    \begin{tabular}{c c *{7}{c}}
        \hline \hline
        \multirow{2}{*}{System} & \multirow{2}{*}{OBMP2} & \multicolumn{3}{c}{O2BMP2} & \multirow{2}{*}{CC2\cite{loos2023heptazine}} & \multirow{2}{*}{SOS-CC2\cite{loos2023heptazine}} & \multirow{2}{*}{ADC(2)\cite{loos2023heptazine}} & \multirow{2}{*}{SOS-ADC(2)\cite{loos2023heptazine}} \\
        \cline{3-5}
         & & {1.2} & {1.5} & {1.8} & & & & \\
        \hline
        1 & 0.557 & -0.037 & -0.053 & -0.286 & -0.239 & -0.389 & -0.246 & -0.452 \\
        2 & 0.292 & -0.082 & -0.127 & -0.235 & -0.130 & -0.270 & -0.137 & -0.273 \\
        3 & 0.194 & -0.114 & -0.163 & -0.201 & -0.106 & -0.241 & -0.113 & -0.246 \\
        4 & 0.073 & -0.115 & -0.153 & -0.186 & -0.131 & -0.275 & -0.139 & -0.278 \\
        5 & -0.007 & -0.079 & -0.091 & -0.127 & -0.118 & -0.257 & -0.127 & -0.263 \\
        6 & 0.192 & -0.196 & -0.152 & -0.165 & -0.085 & -0.213 & -0.094 & -0.218 \\
        7 & 0.121 & -0.035 & -0.129 & -0.211 & -0.065 & -0.191 & -0.074 & -0.196 \\
        8 & 0.431 & -0.001 & -0.043 & -0.154 & -0.058 & -0.185 & -0.070 & -0.191 \\
        9 & -0.016 & -0.189 & -0.223 & -0.389 & -0.212 & -0.367 & -0.214 & -0.633 \\
        10 & 0.409 & -0.126 & -0.214 & -0.349 & -0.446 & -0.517 & -0.435 & -0.503 \\
        \hline \hline
    \end{tabular}
    \label{tab:cos}
\end{table*}

\begin{table}[h]
    \centering
    \caption{$\Delta E_{\text{ST}}$ for 10 INVEST molecules adapted from Ref.~\citenum{loos2023heptazine} using O2BMP2/1.2 with different BCH orders.}
    \label{tab:bch}
    \begin{tabular}{ccccccc}
        \hline \hline
        \multirow{2}{*}{System}&& \multicolumn{2}{c}{$1^{\text{st}}$ BCH order}
          && \multicolumn{2}{c}{$2^{\text{nd}}$ BCH order}\\
        \cline{3-4} \cline{6-7}
        && $1^{\text{st}}$ iter & converged && $1^{\text{st}}$ iter &converged \\
        \hline
        1 & & 1.320 & 0.263  && 0.598  & -0.037  \\
        2 & & 0.800 & 0.057  && 0.302  & -0.082  \\
        3 & & 0.429 & -0.114 && 0.080  & -0.114 \\
        4 & & 0.578 & 0.037  && 0.179  & -0.115  \\
        5 & & 0.588 & 0.014  && 0.211  & -0.079  \\
        6 & & 0.249 & -0.156 && -0.027 & -0.196 \\
        7 & & 0.526 & 0.037  && 0.175  & -0.035  \\
        8 & & 0.875 & 0.114  && 0.397  & -0.001  \\
        9 & & 0.312 & -0.161 && -0.028 & -0.189 \\
        10 & & 1.401 & 0.026 && 0.596  & -0.126  \\
        \hline \hline
    \end{tabular}
\end{table}

\begin{figure*}[t!]
  \includegraphics[width=12cm]{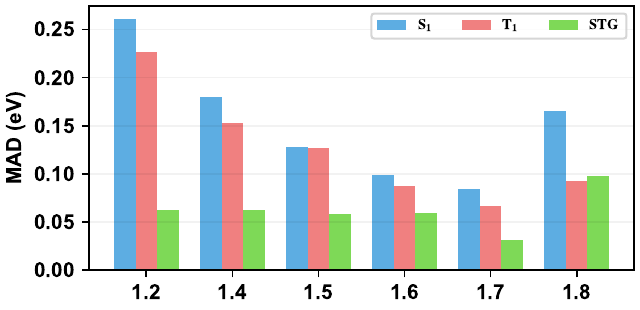}
  \caption{Mean absolute deviations (MADs) of the lowest singlet and triplet excitation energies and the resulting singlet–triplet energy gap}
  \label{fig:mae_tbe}
\end{figure*}

We now assess the accuracy of O2BMP2 for predicting the singlet–triplet energy gap, $\Delta E_{\text{ST}}$. The theoretical best estimates (TBEs) reported in Ref~\citenum{loos2023heptazine} are used as references. To select an optimal spin-opposite scaling factor, we display in Figure~\ref{fig:mae_tbe} the mean absolute deviations (MADs) for the lowest singlet and triplet excitation energies, $\Delta E_{\text{S1}}$ and $\Delta E_{\text{T1}}$, respectively, and for the resulting $\Delta E_{\text{ST}}$, as a function of $c_{\text{os}}$ in the range 1.2-1.8. Generally, the MAD for T1 is smaller than that for S1. For $c_{\text{os}}$ between 1.2 and 1.6, the MADs of $\Delta E_{\text{S1}}$ and $\Delta E_{\text{T1}}$ decrease appreciably with increasing $c_{\text{os}}$, while the MAD of $\Delta E_{\text{ST}}$ remains near 0.062 eV, which is comparatively sizable. At $c_{\text{os}} = 1.7$, O2BMP2 exhibits the best agreement with TBEs for $\Delta E_{\text{S1}}$, $\Delta E_{\text{T1}}$, and $\Delta E_{\text{ST}}$. Accordingly, we adopt $c_{\text{os}} = 1.7$ as the optimal value for subsequent comparisons.

\begin{table*}[h]
    \centering
    \caption{Errors of $\Delta E_{\text{ST}}$ obtained from different methods relative to TBE\cite{loos2023heptazine} for the INVEST set A. For O2BMP2, the spin-opposite scaling $c_{\text{os}} = 1.7$ is used.
    }
    \label{tab:errors_tbe}
    \begin{tabular}{c *{5}{c}}
        \hline \hline
        {System} & {$\Delta$HF} &  {$\Delta$MP2}&  {$\Delta$CCSD}\cite{kunze2024deltadft} &  {$\Delta$CCSD(T)}\cite{kunze2024deltadft}&  {O2BMP2/1.7} \\
        \hline
        1 & -2.071 & 2.456 & -0.260 & 0.037 & 0.034  \\
        2 & -1.559 & 1.623 & -0.251 & 0.030 & -0.026  \\
        3 & -1.215 & 0.918 & -0.206 & 0.033 & -0.067  \\
        4 & -1.294 & 1.153 & -0.207 & 0.051 & -0.010 \\
        5 & -1.194 & 1.120 & -0.169 & 0.065 & -0.007  \\
        6 & -1.063 & 0.652 & -0.207 & 0.018 & -0.042  \\
        7 & -1.185 & 0.952 & -0.221 & 0.021 & -0.057  \\
        8 & -1.486 & 1.573 & -0.292 & -0.023 & -0.025  \\
        9 & -1.189 & 0.918 & -0.150 & 0.081 & 0.005  \\
        10 & -2.318 & 2.669 & -0.287 & -0.033 & 0.041 \\
        \hline \hline
         {MAD} & 1.457 & 1.403 & 0.225 & 0.039 & 0.031\\
         {STD} & 0.420 & 0.681 & 0.020 & 0.020 & 0.021 \\
        \hline
    \end{tabular}\\
\end{table*}

Table \ref{tab:errors_tbe} lists the errors relative to TBEs for O2BMP2/1.7, as well as $\Delta$HF and $\Delta$MP2. All underlying data are provided in the SI. For comparison, we also include the $\Delta$CCSD and $\Delta$CCSD(T) results reported in Ref\citenum{kunze2024deltadft}. With the exception of $\Delta$MP2, all other methods yield inversion of the singlet–triplet energy gaps, but their quantitative accuracy varies considerably across methods and levels of theory. Specifically, $\Delta$HF yields markedly negative $\Delta E_{\text{ST}}$ due to excessive orbital relaxation in the absence of correlation, producing unphysical gaps \cite{won2023inverted,wrigley2025singlet}. $\Delta$MP2 reduces the MAD to 0.792 eV, but remains relatively large and fails to predict the gap inversion. $\Delta$CCSD properly captures the inversion and reduces the MAD to 0.225 eV, which is still notably large. Incorporating perturbative triples in $\Delta$CCSD(T) significantly lowers the MAD to 0.039 eV. As shown in Table~\ref{tab:errors_tbe}, O2BMP2/1.7, with a formal $N^5$ scaling, attains a chemical-accuracy MAD of 0.031 eV, outperforming the higher-cost $\Delta$CCSD ($N^6$) and approaching the accuracy of $\Delta$CCSD(T) ($N^7$). Overall, O2BMP2/1.7 demonstrates a reliable performance for INVEST molecules, attributable to the synergy of self-consistency, the spin-opposite component, appropriate electron correlation, and explicit treatment of double excitations.
        
\subsection{INVEST SET B}

We now evaluate the efficacy of the O2BMP2/1.7 method using a test set of 20 medium-sized molecules selected from the dataset reported by Pollice and co-workers\cite{pollice2024rational}. This dataset comprises a substantial collection of molecules that simultaneously exhibit inverted singlet–triplet energy gaps and appreciable fluorescence rates. The authors employed a hierarchy of excited-state methods capable of capturing double excitations, specifically ADC(2), ADC(3), EOM-CCSD, FNO-EOM-CCSD, and NEVPT2, thereby establishing a robust reference dataset for INVEST molecules. For comparative analysis, we have adopted the ADC(2), ADC(3), and EOM-CCSD results from their study. Additionally, we performed $\Delta$DFT calculations utilizing the CAM-B3LYP and PBE0 functionals. 

\begin{figure*}[t!]
  \includegraphics[width=15cm]{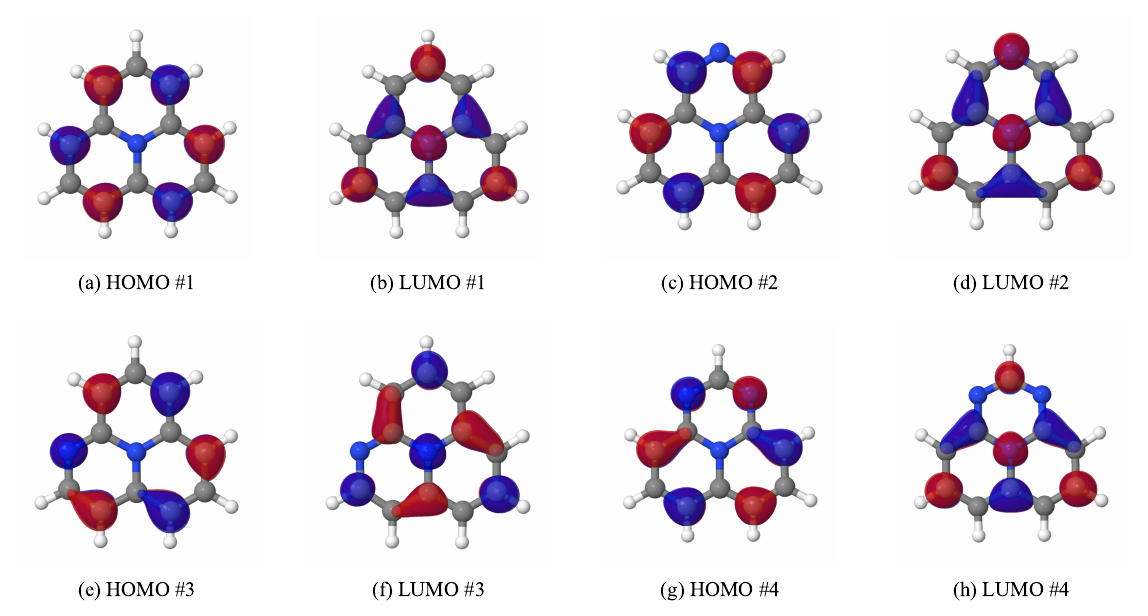}
  \caption{HOMO and LUMO of first four molecules in the INVEST set B calculated using O2BMP2/1.7.}
  \label{fig:HOMO-LUMO}
\end{figure*}

The magnitude of $\Delta E_{\text{ST}}$ is directly governed by the exchange integral between the highest occupied molecular orbital (HOMO) and the lowest unoccupied molecular orbital (LUMO). Minimal overlap between these frontier orbitals is a prerequisite for singlet-triplet gap inversion. Figure~\ref{fig:HOMO-LUMO} displays the HOMO and LUMO for the first four molecules in the INVEST set B, calculated using O2BMP2/1.7. It is evident that the HOMO and LUMO possess spatially distinct distributions. The HOMO is delocalized over the conjugated framework, exhibiting nodal features that minimize exchange stabilization, thereby favoring a higher-lying triplet state. Conversely, the LUMO shows complementary localization, with lobe amplitudes concentrated on alternating rings or N regions. This results in minimal overlap with the HOMO, significantly reducing the exchange integral. Such spatial separation stabilizes the triplet configuration relative to the singlet, ultimately yielding an inverted gap.

%\begin{figure}[h!]
%  \centering
%  \captionsetup[sub]{font=normalsize}
%  \resizebox{1.0\textwidth}{!}{%
%    \begin{tabular}{cccc}
%      \subcaptionbox{HOMO \#1\label{fig:a}}%
%      {\includegraphics[width=0.23\textwidth]{alan-orbital/alan-1-17-44.png}} &
%      \subcaptionbox{LUMO \#1\label{fig:b}}%
%      {\includegraphics[width=0.23\textwidth]{alan-orbital/alan-1-17-45.png}} &
%      \subcaptionbox{HOMO \#2\label{fig:c}}%
%      {\includegraphics[width=0.23\textwidth]{alan-orbital/alan-2-17-44.png}} &
%      \subcaptionbox{LUMO \#2\label{fig:d}}%
%      {\includegraphics[width=0.23\textwidth]{alan-orbital/alan-2-17-45.png}} \\
%
%      \subcaptionbox{HOMO \#3\label{fig:e}}%
%      {\includegraphics[width=0.23\textwidth]{alan-orbital/alan-3-17-44.png}} &
%      \subcaptionbox{LUMO \#3\label{fig:f}}%
%      {\includegraphics[width=0.23\textwidth]{alan-orbital/alan-3-17-45.png}} &
%      \subcaptionbox{HOMO \#4\label{fig:g}}%
%      {\includegraphics[width=0.23\textwidth]{alan-orbital/alan-4-17-44.png}} &
%      \subcaptionbox{LUMO \#4\label{fig:h}}%
%      {\includegraphics[width=0.23\textwidth]{alan-orbital/alan-4-17-45.png}} \\
%    \end{tabular}
%  }
%  \caption{HOMO and LUMO of first four molecules in the INVEST set B calculated using O2BMP2/1.7.}
%  \label{fig:HOMO-LUMO}
%\end{figure}

\begin{figure*}[h!]
    \includegraphics[width=12cm]{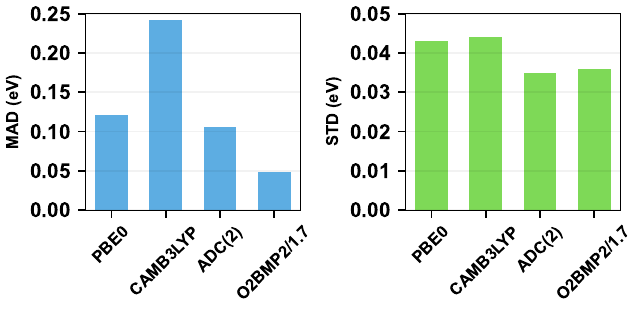}
    \includegraphics[width=12cm]{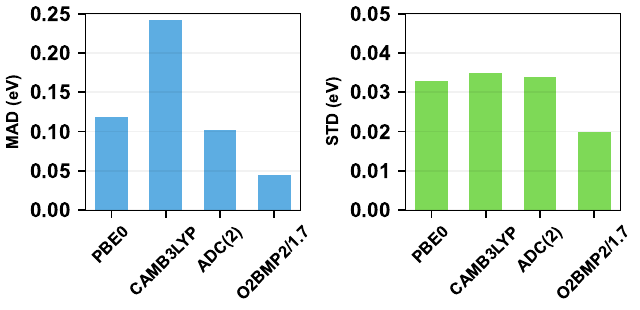}
    \caption{MADs and STDs relative to EOM-CCSD (upper panel) and ADC(3) (lower panel).}
    \label{fig:mae_std_alan}
\end{figure*}

All $\Delta E_{\text{ST}}$ values of the INVEST set B are listed in Table \ref{tab:stgap_alan}, and statistical errors relative to EOM-CCSD and ADC(3) are illustrated in Figure \ref{fig:mae_std_alan}. Although ADC(2) uniformly produces negative $\Delta E_{\text{ST}}$ values, it tends to overestimate the magnitude of the inversion compared to ADC(3) and EOM-CCSD, which predict slightly positive gaps for several molecules. Similarly, $\Delta$DFT (using PBE0 and CAM-B3LYP) frequently yields negative $\Delta E_{\text{ST}}$ values but largely underestimates them relative to the wavefunction references. O2BMP2/1.7 delivers negative $\Delta E_{\text{ST}}$ values for most systems, excluding molecules 14 and 20, where the gaps are slightly positive and near zero. The trends in MAD are consistent across both references, with small STDs indicating systematic error behavior. While CAM-B3LYP exhibits significant deviations, PBE0 achieves accuracy comparable to ADC(2). Notably, O2BMP2/1.7 reduces the MADs by approximately 50\% compared to PBE0 and ADC(2). In conclusion, our method offers an appropriate description of INVEST excited states; it identifies inverted gaps in more systems than ADC(3) or EOM-CCSD, yet maintains closer quantitative agreement with these high-level references than either $\Delta$DFT or ADC(2).

\begin{table*}[h]
    \centering
    \caption{$\Delta E_{\text{ST}}$ for 20 medium-size molecules selected from Ref~\citenum{pollice2024rational} using various methods}
    \label{tab:stgap_alan}
    \begin{tabular}{c *{6}{c}}
        \hline \hline
         {Molecule} &  {O2BMP2} &  {PBE0} &  {CAM-B3LYP} &  {ADC(2)}\cite{pollice2024rational} &  {ADC(3)}\cite{pollice2024rational} &  {EOM-CCSD} \cite{pollice2024rational}\\
        \hline
        1  & -0.163 & -0.179 & -0.328 & -0.160 & -0.093 & -0.099 \\
        2  & -0.106 & -0.139 & -0.262 & -0.124 & -0.046 & -0.047 \\
        3  & -0.108 & -0.156 & -0.291 & -0.111 & -0.043 & -0.022 \\
        4  & -0.106 & -0.159 & -0.290 & -0.148 & -0.054 & -0.068 \\
        5  & -0.037 & -0.152 & -0.290 & -0.139 & -0.017 & -0.034 \\
        6  & -0.060 & -0.119 & -0.230 & -0.066 & 0.005 &  0.042 \\
        7  & -0.084 &  0.000 & -0.278 & -0.149 & -0.043 & -0.060 \\
        8  & -0.112 & -0.158 & -0.281 & -0.101 & -0.051 & -0.068 \\
        9  & -0.090 & -0.167 & -0.292 & -0.158 & -0.043 & -0.061 \\
        10 & -0.068 & -0.139 & -0.253 & -0.128 & -0.014 & -0.078 \\
        11 & -0.076 & -0.153 & -0.260 & -0.171 & -0.027 & -0.071 \\
        12 & -0.018 & -0.104 & -0.199 & -0.059 & -0.045 & 0.063 \\
        13 & -0.002 & -0.125 & -0.235 & -0.063 & 0.053 & 0.076 \\
        14 & -0.019 & -0.149 & -0.260 & -0.131 & 0.002 &  -0.013 \\
        15 & -0.026 & -0.177 & -0.322 & -0.188 & -0.028 &  0.028 \\
        16 & -0.015 & - & - & -0.066 & -0.061 &  0.124 \\
        17 & -0.036 & -0.082 & -0.160 & -0.080 & 0.033 &  0.026 \\
        18 & -0.039 & -0.098 & -0.189 & -0.092 & 0.016 &  0.011 \\
        19 & -0.005 & -0.130 & -0.264 & -0.113 & 0.013 &  0.008 \\
        20 &  0.014 & -0.107 & -0.219 & -0.088 & 0.045 &  0.045 \\
        \hline \hline
    \end{tabular}
    \\
\end{table*}

\begin{figure*}[h!]
  \includegraphics[width=\linewidth]{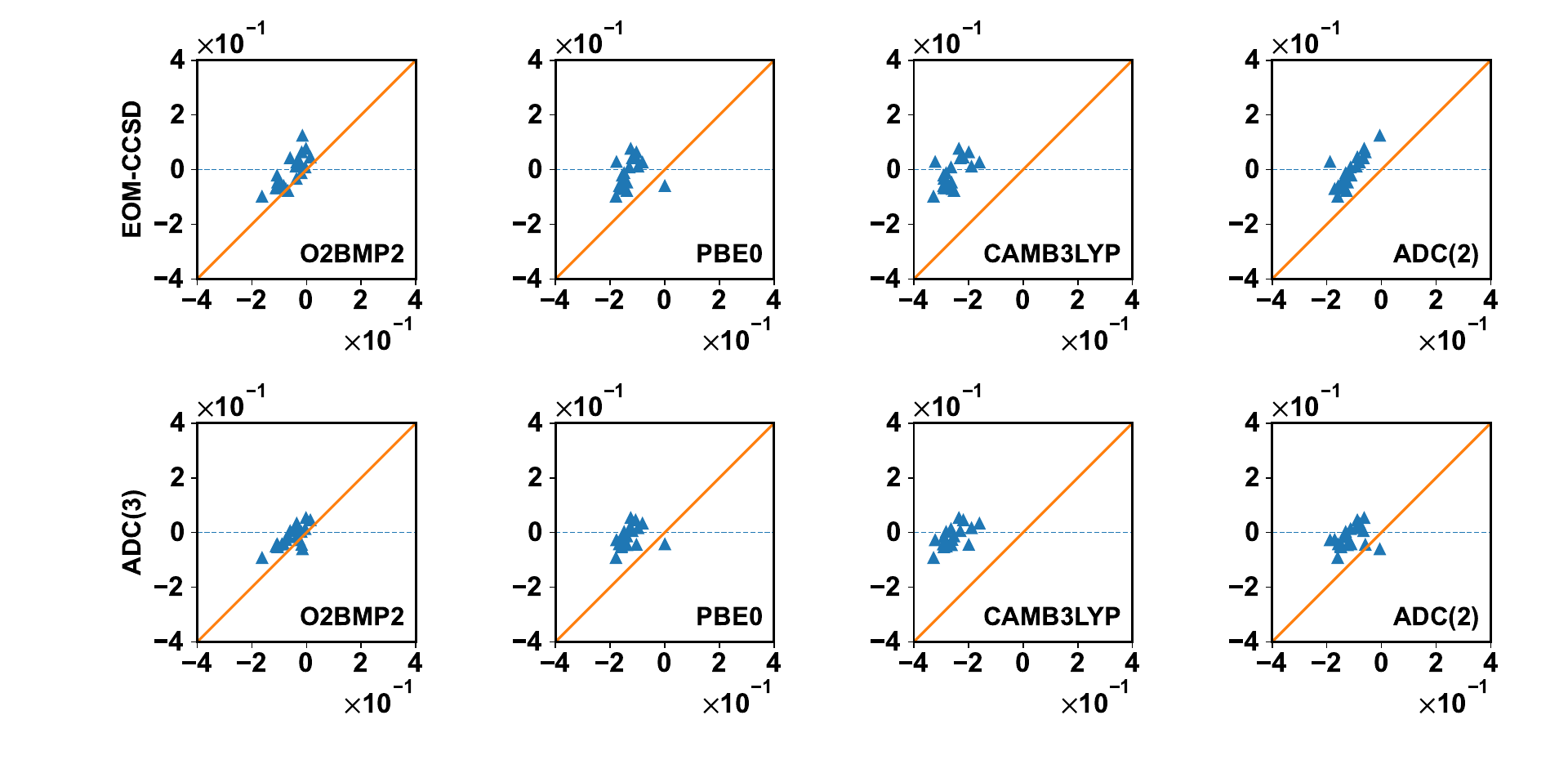}
  \caption{Pearson correlation between different methods with EOM-CCSD (upper panel) and ADC(3) (lower panel).}
  \label{fig:pearson}
\end{figure*}

Recently, Valverde and co-workers\cite{valverde2025can} employed the Pearson correlation coefficient of DFT with respect to EOM-CCSD to evaluate the method's applicability in high-throughput and machine-learning workflows for identifying INVEST materials. The Pearson correlation quantifies the strength and direction of the linear relationship between two datasets. In this work, we computed the Pearson correlation coefficients of O2BMP2 with respect to both EOM-CCSD and ADC(3) for test set B to determine whether O2BMP2 can serve as a reliable method for screening applications. The results are summarized in Figure \ref{fig:pearson} and Table \ref{tab:pearson}. Although PBE0 was previously found to be the best functional for the $\Delta$DFT approach in predicting INVEST molecules, it exhibits a weak correlation with EOM-CCSD and ADC(3), yielding low coefficients of 0.32 and 0.27, respectively. Surprisingly, while CAM-B3LYP produces larger MADs than PBE0, it demonstrates a stronger correlation with both references. Furthermore, while ADC(2) shows a strong correlation with EOM-CCSD (0.82), it correlates poorly with ADC(3) (0.35). This discrepancy underscores the challenges associated with accurately describing inverted singlet–triplet gaps, even within high-level correlated methods. In contrast, O2BMP2 exhibits a remarkably consistent correlation with both EOM-CCSD and ADC(3), achieving coefficients of 0.79 and 0.73, respectively. This consistency indicates the capability of O2BMP2 to capture the qualitative trends of ST gap inversion, thereby demonstrating its strong potential for reliable and efficient INVEST screening. 

\begin{table}[h]
    \centering
    \caption{Pearson correlation coefficients between various methods with respect to ADC(3) and EOM-CCSD references.}
    \label{tab:pearson}
    \begin{tabular}{c *{6}{c}}
        \hline \hline
         {References} &  {O2BMP2} &  {PBE0} &  {CAM-B3LYP} &  {ADC(2)}\\
        \hline
        ADC(3)    & 0.73 & 0.32 & 0.65 & 0.35 \\
        EOM-CCSD  & 0.79 & 0.27 & 0.60 & 0.82 \\
        \hline \hline
    \end{tabular}
\end{table}

\begin{figure*}[h!]
  \includegraphics[width=15cm]{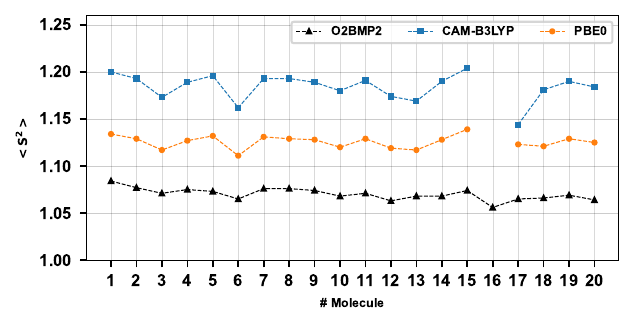}
  \caption{$\left<S^2\right>$ values for the first singlet excited state (S1) of 20 molecules in the INVEST set B.}
  \label{fig:spin-s1}
\end{figure*}

One of the primary drawbacks of the $\Delta$SCF approach is its susceptibility to spin contamination, which arises from the use of a single-determinant wavefunction to describe excited singlet states. In particular, the singlet state obtained via unrestricted SCF (UKS, UHF, and UO2BMP2) is generally a 50:50 mixture of singlet and triplet states, characterized by an ideal expectation value of $\left<S^2\right> = 1.00$\cite{yamaguchi1988spin,kitagawa2007approximately,valverde2025can,kunze2024deltadft}. In Figure~\ref{fig:spin-s1}, we plot the $\left<S^2\right>$ values for the S1 state of 20 molecules in the INVEST set B obtained from $\Delta$SCF calculations using PBE0, CAM-B3LYP, and O2BMP2/1.7. It is evident that PBE0 exhibits significantly less spin contamination than CAM-B3LYP, explaining its superior performance relative to higher-level methods, as shown in Figure~\ref{fig:mae_std_alan}. Notably, O2BMP2 yields the lowest spin contamination and close to the ideal value. Overall, this spin contamination analysis further corroborates the reliability of O2BMP2 for predicting INVEST properties. 

\section{Conclusion}

In conclusion, we assessed the performance of our methods, OBMP2 and O2BMP2, on 30 INVEST molecules divided into two test sets: INVEST A and B. We found that while OBMP2 fails to describe INVEST excited states with positive singlet-triplet energy gaps, O2BMP2 successfully captures the inversion of energy gaps by accounting for double excitations, charge transfer, and stronger electron correlation effects relative to exchange. Notably, with an appropriate spin-opposite scaling of $c_{\text{os}}$ = 1.7, O2BMP2 yields $\Delta E_{\text{ST}}$ values with an accuracy comparable to ADC(3) and EOM-CCSD. Further analyses of molecular frontier orbitals and spin contamination corroborate the reliability of O2BMP2 for predicting INVEST properties. Crucially, as the formal scaling of O2BMP2 can be reduced from $N^5$ to $N^4$, the method offers high accuracy at an affordable computational cost. Consequently, O2BMP2 is well-suited for high-throughput screening aimed at identifying new molecular systems exhibiting INVEST excited states.

\section*{Acknowledgment}
This research is funded by the National Foundation for Science and Technology Development (NAFOSTED) Grant Number 103.01-2024.06.

\section*{Supplementary Information}
Raw data for S1 and T1 excitation energies of all molecules (in eV).

%\subsection*{Conflict of Interest}
%The authors have no conflicts to disclose.

\bibliography{main}
\end{document}